# Two-phase immiscible displacement of water by gas accounting for fines migration: analytical model for vertical equilibrium

Amin Shokrollahi, Kofi Ohemeng Kyei Prempeh, Syeda Sara Mobasher, Abbas Zeinijahromi, Pavel Bedrikovetsky

*School of Chemical Engineering, Discipline of Mining and Petroleum Engineering, The University of Adelaide, Adelaide, SA 5005, Australia*

**Key points**

- Extending the capillary-gravity equilibrium model to include fines migration.
- Deriving analytical solutions for one-dimensional radial flow with fines migration.
- Conducting sensitivity analysis on reservoir heterogeneity, gas viscosity, and damage.
- Highlighting fines migration's dual impact on well injectivity and sweep efficiency.

**Abstract**

Fines migration during subterranean $CO_2$ storage is considered as one of major contributors to formation damage and injectivity decline. The main formation damage mechanism by migrating fines is rock clogging by straining of particles in thin pore throats. The main mobilisation mechanism of attached natural reservoir fines is their detachment by capillary forces exerting from the advancing water-gas menisci. The mathematical model for displacement of water by injected $CO_2$ from layer-cake reservoirs under fines mobilisation, migration and straining extends the classical vertical capillary-gravity equilibrium model, where hydrostatic pressure gradient is assumed in each phase. The analytical model generalises classical Buckley-Leverett problem, where the extended fractional flow accounts for permeability damage due to fines migration. We show the significant effect of fines migration on water displacement – while it decreases well injectivity, fines migration simultaneously increases reservoir sweep, increasing $CO_2$ storage capacity.

## Nomenclature

**Parameters**

| | |
|---|---|
| $B$ | Power law exponent for MRF, [-] |
| $f(S)$ | Fractional flow of gas, [-] |
| $f_0$ | Fractional flow in Zone 0, [-] |
| $f_{01}$ | Fractional flow in Zone 01, [-] |
| $f_1$ | Fractional flow in Zone 1, [-] |
| $g$ | Gravitational acceleration, [LT$^{-2}$] |
| $H$ | Total height of reservoir, [L] |
| $h$ | Height of gas, [L] |
| $h_c$ | Dimensionless height of transition zone, [-] |
| $J(s)$ | Leverett J function, [-] |
| $J(T)$ | Well Impedance, [-] |
| $j$ | Invers of Leverett function, [-] |
| $K_{rgwc}$ | Gas relative permeability at irreducible water saturation, [-] |
| $k_{rw}$ | Water relative permeability, [-] |
| $k(z)$ | Horizontal permeability as a function of depth, [L$^2$] |
| $k(Z)$ | Horizontal permeability as a function of dimensionless depth, [L$^2$] |
| $K(Z)$ | Dimensionless horizontal permeability as a function of dimensionless depth, [-] |
| $k_{min}$ | Minimum permeability, [L$^2$] |
| $k_{max}$ | Maximum permeability, [L$^2$] |
| $k_0$ | Average permeability, [L$^2$] |
| $p$ | Pressure, [ML$^{-1}$T$^{-2}$] |
| $P_c$ | Capillary pressure [ML$^{-1}$T$^{-2}$] |
| $P_{c_D}$ | Dimensionless capillary pressure, [-] |
| $P_d$ | Capillary threshold pressure [ML$^{-1}$T$^{-2}$] |
| $P_m$ | Maximum capillary pressure [ML$^{-1}$T$^{-2}$] |
| $P$ | Dimensionless pressure, [-] |
| $Q$ | Rate, [L$^3$T$^{-1}$] |
| $R$ | Drainage radius, [L] |
| $s$ | Gas saturation, [-] |
| $S$ | Normalised gas saturation, [-] |
| $S_0$ | Gas saturation in Zone 0, [-] |
| $S_{01}$ | Gas saturation in Zone 01, [-] |
| $S_1$ | Gas saturation in Zone 1, [-] |
| $S_{wc}$ | Irreducible water saturation, [-] |
| $S^*$ | Normalise gas saturation at the top of the reservoir in region 0 of thick reservoirs, [-] |
| $S^{**}$ | Normalise gas saturation at the top of the reservoir in region 0 of thin reservoirs, [-] |
| $S_*$ | Normalise gas saturation at the bottom of the reservoir in region 1 of thin and thick reservoirs, [-] |
| $t$ | Time, [T] |
| $T$ | Dimensionless time [-] |
| $U_g$ | Gas flux, [L$^2$T$^{-1}$] |
| $U_w$ | Water flux, [L$^2$T$^{-1}$] |
| $u_g$ | Radial gas velocity, [LT$^{-1}$] |
| $u_w$ | Radial water velocity, [LT$^{-1}$] |
| $x$ | Dimensionless distance, [-] |
| $z$ | Depth, [L] |
| $Z$ | Dimensionless depth, [-] |
| $z_0$ | Position of the advanced front, [L] |
| $z_1$ | Position of the receded front, [L] |
| $\Delta P$ | Dimensionless pressure drops, [-] |

*Greek Symbols*

| | |
|---|---|
| $\beta$ | Formation damage coefficient, [-] |
| $\phi$ | Porosity, [-] |
| $\mu_w$ | Water viscosity, [ML$^{-1}$T$^{-1}$] |
| $\mu_g$ | Gas viscosity, [ML$^{-1}$T$^{-1}$] |
| $\rho_w$ | Water density, [ML$^{-3}$] |
| $\rho_g$ | Gas density, [ML$^{-3}$] |
| $\Lambda(S)$ | Total mobility, [-] |
| $\xi$ | Self-similar variable, [-] |
| $\sigma_0$ | Initial particle concentration, [-] |
| $\sigma_s$ | Strained particle concentration, [-] |
| $\gamma$ | Dimensionless parameter, [-] |
| $\lambda$ | Corey coefficient, [-] |

*Abbreviations*

| | |
|---|---|
| CCS | Carbon capture and storage |
| MRF | Maximum retention function |

# 1 Introduction

Carbon capture and storage (CCS) is an emerging technological approach with the potential to mitigate atmospheric $CO_2$ levels. To significantly address the climate crisis, the implementation of CCS must occur on a vast scale, capable of managing a substantial portion of the growing carbon emissions [1]. Injecting $CO_2$ into deep saline aquifers offers a significant capacity for $CO_2$ storage. Assessing the potential risks of $CO_2$ leakage from these aquifers, as well as the behaviour of the injected $CO_2$ over both short and long-term periods, requires simulations that incorporate extensive temporal and spatial scales. Given the inherent uncertainties in geological assessments, it will be essential to simulate various possible outcomes for each storage scenario to evaluate the associated risks. This highlights the need for the development of rapid simulation tools.

In recent years, there has been a resurgence of interest in Vertical Equilibrium (VE) methods as a tool for simulating large-scale $CO_2$ migration, where the assumption of sharp interfaces and vertical equilibrium may be appropriate. Several researchers have developed analytical solutions to explore different aspects of $CO_2$ injection, assuming rapid vertical segregation and vertical equilibrium [2-8]. Notably, Gasda et al. [9] extended a VE formulation using sub-scale analytic functions, demonstrating the potential of VE models to accelerate $CO_2$ migration simulations. Numerical results obtained with the VE formulation showed good agreement with full 3D simulations in a recent benchmark study [10].

Displacement processes in oil reservoirs or water aquifers are often simplified when the reservoir is narrow and elongated, and the flow is nearly parallel, a scenario common in many applications. Researchers have proposed various approximations for such conditions. Typically, a VE assumption is made. Based on the influence of gravity, these approaches can be divided into two categories: one where viscous forces and heterogeneity dominate the phase distribution, and another where gravity causes complete phase segregation [11].

The first category focuses primarily on the effects of viscous forces and their interaction with heterogeneity. This has been examined by several researchers, such as Coats et al. [12], Yokoyama and Lake [13], Zapata and Lake [14], and Pande and Orr [15]. A comprehensive discussion of VE based on physical principles can be found in Lake [16]. In this context, the term "vertical" refers to the direction along the transverse coordinate. In the majority of these studies, a two-layer model is employed, which, although based on intuitive reasoning, has proven to be correct upon later examination. Extensive numerical simulations have confirmed the validity of these approaches, particularly in relation to the dimensionless parameter $R_L = L/H \sqrt{k_v/k_H}$, which must attain sufficiently large values for VE to be applicable. Similarly, the work of Lake and Hirasaki [17] on tracer dispersion in stratified systems, as well as various phenomenological models of viscous fingering, such as those by Koval [18], Todd and Longstaff [19], and Fayers [20], merit consideration. Although these models are primarily empirical, numerical evidence often supports their relevance and applicability in such contexts [11].

The second category focuses on the combined effects of gravity and viscous forces, making it particularly relevant to systems with higher permeability. The initial developments in this area were related to groundwater aquifers, where the Dupuit assumption was first introduced [21]. The study of viscous, two-phase flow was conducted by Dietz [22] and further developed by Le Fur and Sourieau [23], Beckers [24], and others. This approach assumes complete segregation of the immiscible phases, with a distinct macroscopic interface separating the two regions. Fayers and Muggeridge [25] then extended this framework to tilted reservoirs with a dip.

Although the two categories appear to stem from similar conditions, no systematic effort has been made to treat them uniformly. In fact, it remains unclear which parameters in the parameter space define the boundaries between the two regimes and where each approximation is valid. Currently, most of the available evidence is numerical, which may be adequate in some cases. However, a rigorous derivation is still needed to clearly identify the various approximations and assumptions. This is especially important for layered systems, where current models are often complex and difficult to apply to multiple layers.

A crucial factor in the effective storage of $CO_2$ within subsurface aquifers is the injectivity of wells, which plays a central role in determining the success of $CO_2$ injection into geological formations. Formation damage, however, can notably obstruct the flow of fluids through the porous media, thereby reducing injectivity. This restriction can have a detrimental impact on the efficiency of $CO_2$ storage operations, ultimately affecting the economic feasibility of such projects [26-28].

Fines migration is a widely recognised and detrimental cause of formation damage [29, 30]. It occurs when natural fines, which are initially attached to the internal surfaces of reservoir rocks, become detached and are subsequently mobilised. As these fines migrate through the pore network, they can become trapped in narrow pore throats, leading to a significant reduction in the permeability of the rock. This decline in permeability can severely hinder fluid flow. The impact of fines migration on well productivity, particularly before water breakthrough, is typically described using single-phase flow equations in porous media, which consider the migration of fines [31].

The presence of multiple phases in the porous media introduces additional complexity to the system. In two-phase flow, models for particle detachment must consider the influence of residual phases, the effect of each flowing phase, and the role of the fluid–fluid interface. These factors collectively impact the behaviour of particles within the porous medium [32-35].

During gas injection, an increase in gas saturation leads to the displacement of water by gas through the gas–water menisci in intermediate-sized capillaries. This process results in the detachment of fines due to capillary forces. The detached fines are carried by the menisci, although some fines become gradually trapped in narrower pores. These strained fines remain within the pores that are saturated by the displacing phase, which reduces the permeability of the phase. Gas flow predominantly occurs through the largest pores. When gas velocity is high, the drag force may detach additional particles, which can then migrate further within the gas zone [36].

In two-phase flow, the fluid–fluid interface exerts an additional capillary force on certain particles that are attached to the porous media [37]. Experimental and modelling studies have demonstrated that this capillary force can play a dominant role in two-phase flows, particularly in porous media and narrow channels [32-35]. As the fluid–fluid meniscus moves across a particle, it applies a force that can either attach or detach the particle. The magnitude and direction of this force are significantly influenced by factors such as the fluid–fluid-particle contact angle, the fluid–fluid-rock contact angle, and the direction of meniscus movement, whether in drainage or imbibition.

Critical or maximum retention function (MRF) is used to quantify the fines detachment due to capillary forces [37]. The MRF enables the closure of the governing system for fines migration at core and above scales, specifically during single-phase colloidal-suspension-nano transport in porous media [38]. Analytical models for flow that account for fines migration have been developed to assess well productivity [31] and injectivity [39, 40] during single-phase oil or gas production. However, saturation can significantly impact formation damage and well productivity, particularly during commingled oil and water production [41]. To date, no mathematical model has been developed for two-phase flow that accounts for fines detachment via capillary menisci during $CO_2$ injection in a layered aquifer with a layer-cake permeability distribution.

This paper extends the classical vertical capillary-gravity equilibrium model for two-phase flow in layer-cake reservoirs by incorporating the effects of fines migration and the resulting permeability damage. An analytical solution is derived for the one-dimensional problem of radial flow, and a comprehensive sensitivity analysis is conducted to examine the influence of reservoir heterogeneity, gas viscosity, and the formation damage coefficient associated with fines migration.

The manuscript is organised as follows: Section 2 outlines the derivation of the governing equations. Section 3 presents the practical calculations. Section 4 discusses the results from the analytical modelling and performs a sensitivity analysis on the various parameters influencing well injectivity and sweep efficiency. Finally, Section 5 provides a summary and conclusion of the findings.

## 2  Governing equations

This section presents the derivation of the governing equations for $CO_2$ injection into aquifers under vertical capillary-gravity conditions. It begins with the assumptions of the models (Section 2.1), followed by the calculation of depth saturation distribution and fractional flow for both thick and thin reservoirs in Section 2.2. Section 2.3 covers the calculation of well impedance, accounting for water compressibility ahead of the gas front, while Section 2.4 presents the calculation of average gas saturation and sweep efficiency.

## 2.1 Assumptions of the model

The model assumes vertical capillary-gravity equilibrium during $CO_2$ injection into aquifers, considering two immiscible and incompressible phases. At low injection rates, gravity-driven capillary imbibition in the vertical direction is assumed to have sufficient time to attain capillary-gravity equilibrium. Additionally, the model accounts for the low compressibility of water ahead of the advancing $CO_2$ front. These assumptions establish a simplified yet robust framework for capturing the key physical processes governing $CO_2$ migration and storage in aquifers.

## 2.2 Depth saturation distribution

At low injection rates, gravity-driven capillary imbibition in the vertical direction is assumed to have sufficient time to attain capillary-gravity equilibrium. Accordingly, the hydrostatic pressure in each phase is given by:

$$\frac{\partial p_g}{\partial z} = \rho_g g, \quad \frac{\partial p_w}{\partial z} = \rho_w g \tag{1}$$

Water is the wetting phase, so the pressure in the gas phase exceeds that in the water phase by the capillary pressure, expressed as:

$$P_c(s) = p_g - p_w \tag{2}$$

Substituting equation (2) into the second equation in equation (1) and subtracting it from the first equation in equation (1) yields:

$$\frac{dP_c}{dz} = -\Delta\rho g, \quad \Delta\rho = \rho_w - \rho_g \tag{3}$$

Further in the text, the Leverett form of the capillary pressure function, expressed as follows, is used:

$$P_c(s) = \frac{\sigma}{\sqrt{k(z)/\phi}} J(s) \tag{4}$$

Figure 1 illustrates the capillary pressure versus saturation.

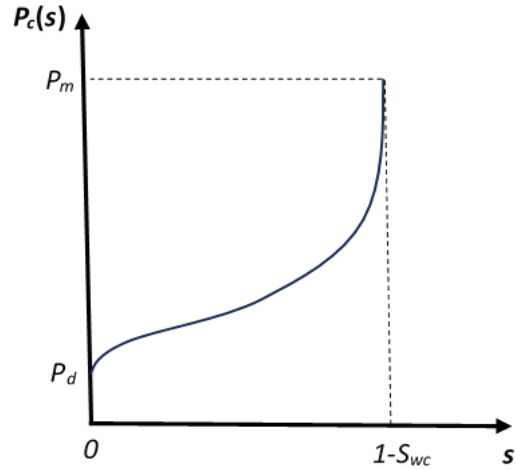

Figure 1. Typical plot of capillary pressure versus saturation.

Integrating equation (3) with respect to $z$ yields:

$$P_c(s) = const - \Delta\rho g z \tag{5}$$

This study considers two cases of thick and thin reservoirs. A thick reservoir is defined as one where the reservoir thickness exceeds that of the transition zone, whereas a thin reservoir is characterised by a reservoir thickness smaller than that of the transition zone. Figure 2 illustrates the two types of reservoirs described above.

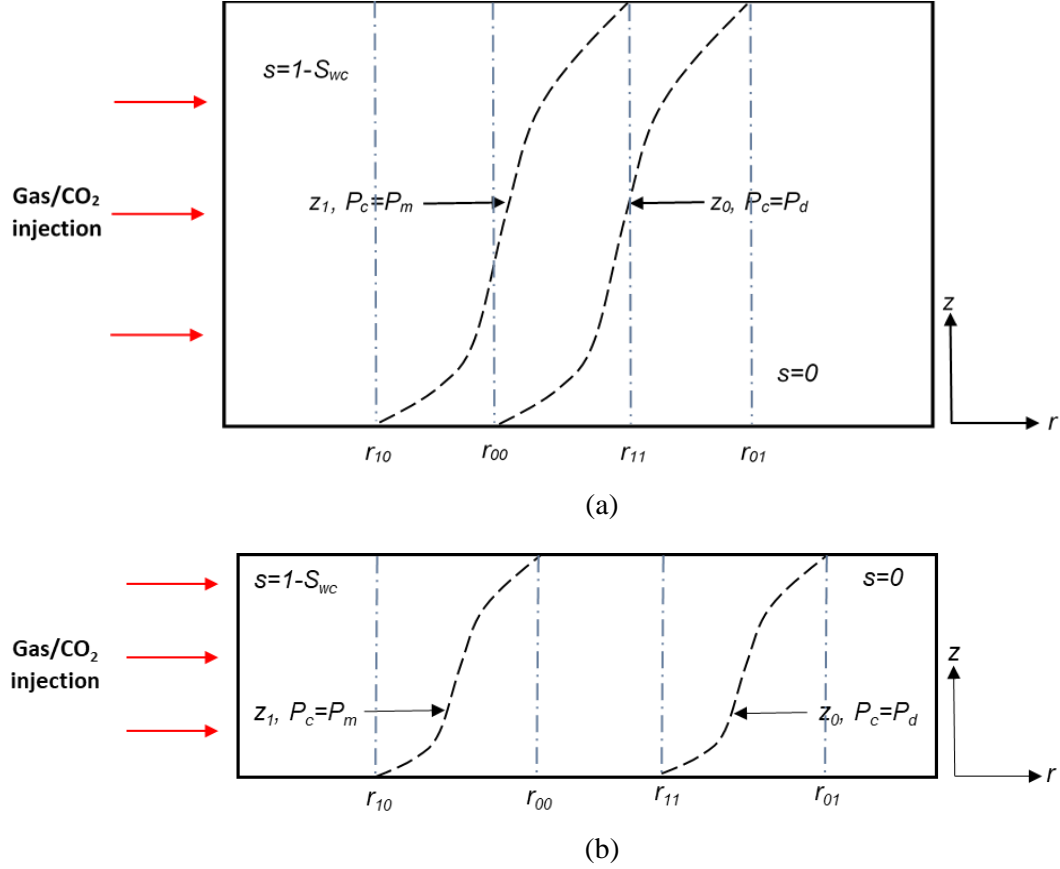

Figure 2. Displacement of water by gas through heterogeneous layer cake reservoir: a) Thick reservoir; b) Thin reservoir.

The advanced water-gas contact $z_0$ and the receded contact $z_1$ are determined as follows:

$$z = z_0 = z_0(x,t): s = 0, P_c = P_d \tag{6}$$

$$z = z_1 = z_1(x,t): s = 1 - S_{wc}, P_c = P_m \tag{7}$$

The constant in equation (5) can be determined using relationships (6) and (7) for the advanced and receded fronts, respectively. This leads to expressions for the capillary pressure.

$$P_c(0) = const - \Delta\rho g z_0, \quad P_c(1 - S_{wc}) = const - \Delta\rho g z_1 \tag{8}$$

Subtracting the first and second equations in equation (8) results in the expression for the thickness of the capillary transition zone, $Hc$.

$$\Delta P_c = P_c(1-S_{wc}) - P_c(0) = -\Delta\rho g(z_1 - z_0) = \Delta\rho g H_c, \quad H > H_c \tag{9}$$

Equation (9) presents the expression for the thickness of the capillary transition zone. Figure 2a illustrates the case for a thick reservoir, where $Hc < H$. In this scenario, the receded front in the layer with maximum permeability reaches every point in the reservoir before the advanced front in the layer with minimum permeability. Figure 2b shows the case of a thin reservoir, where $Hc > H$. Here, the advanced front in the low-permeability layer reaches each reservoir point before the receded front in the highly permeable layer. In the following sections, the governing equations for both thick and thin reservoirs are derived.

### 2.2.1 Governing equation in thick reservoirs ($H > H_c$)

***Advanced and receded fronts***: Consider three cases: Case 0, where the point $x$ is located between the positions of the advanced and receded fronts at $z = H$; Case 01, where $x$ is situated between the position of the receded front at $z = H$ and the advanced front at $z = 0$; and the Case 1 where $x$ lies below the position of the advanced front at $z = 0$ and the receded front at $z = 0$ (Figure 2a).

$$\begin{aligned} 0 &: r_{11} < r < r_{01} \\ 01 &: r_{00} < r < r_{11} \\ 1 &: r_{10} < r < r_{00} \end{aligned} \tag{10}$$

Let us discuss Case 0. Saturation value at the advanced front yields the following depth distributions for capillary pressure and Leverette function $J(s)$ as expressed via the position of advanced front.

$$P_c(s) - P_d = -\Delta\rho g(z - z_0) = \frac{\sigma[J(s) - J_d]}{\sqrt{k(z)/\phi}}, \quad J(s) = J_d - \Delta\rho g(z - z_0)\sqrt{k(z)/\phi}, \quad j = J^{-1} \tag{11}$$

Introducing the inverse $j^{-1}$ to the Leverett function $J(s)$ results in the following depth saturation distribution as a function of the position of the advanced front:

$$s = j\left( J_d - \Delta\rho g(z-z_0)\sqrt{k(z)/\phi} \right) \qquad (12)$$

This is valid for cases 0 and 01 presented in equation (10) (Figure 2a). Using the expression for saturation on the receded front, the depth-dependent saturation yields:

$$P_c(s) - P_m = -\Delta\rho g(z-z_1) = \frac{\sigma[J(s)-J_m]}{\sqrt{k(z)/\phi}}, \quad J(s) = J_m - \Delta\rho g(z-z_1)\sqrt{k(z)/\phi}, \quad j = J^{-1} \qquad (13)$$

The introduction of the inverse function to Leverett's function results in expressing the saturation distribution over the vertical axis as a function of the position of the receded front as follows:

$$s = j\left( J_m - \Delta\rho g(z-z_1)\sqrt{k(z)/\phi} \right) \qquad (14)$$

This is valid for cases 01 and 1 of equation (10) (Figure 2a). For case 01, expressions (12) and (14) are identical.

*Average gas saturation in different regions along the depth*: Figure 2a shows the position $r_{01}$, where the advanced front arrives via the highly permeable layer at time $t$; $r_{11}$, where the receded front arrives via the highly permeable layer at the same moment; and $r_{00}$ and $r_{10}$, which represent the positions of arrival of the advanced and receded fronts via the layer with the lowest permeability at time $t$.

$$z_0(r_{01},t) = H, \quad z_1(r_{11},t) = H, \quad z_0(r_{00},t) = 0, \quad z_1(r_{10},t) = 0 \qquad (15)$$

The average gas saturation is obtained by integrating the gas saturation from equations (12) or (14) over $z$:

$$S(r,z,t) = \frac{1}{H}\int_0^H s(z)\,dz \qquad (16)$$

Particular expressions for average saturation are different for cases 0, 01, and 1.

Saturation ahead of the advanced front is $s=0$, while saturation behind the receded front is equal to $s=1-S_{wc}$. The average saturations over the reservoir thickness for the three cases mentioned above are given by the following formulae:

$$S_0(x,t,z_0) = \frac{1}{H}\int_{z_0}^{H} s(z)dz = \frac{1}{H}\int_{z_0}^{H} j_0\left(J_d - \Delta\rho g(z-z_0)\sqrt{k(z)/\phi}\right)dz \tag{17}$$

$$S_{01}(x,t,z_0) = \frac{1}{H}\int_{z_0}^{z_0+H_c} s(z)dz + \frac{(H-H_c-z_0)(1-S_{wc})}{H} \tag{18}$$

$$S_1(x,t,z_1) = \frac{1}{H}\int_{0}^{H} s(z)dz = \frac{1}{H}\int_{0}^{z_1} j_1\left(J_m - \Delta\rho g(z-z_1)\sqrt{k(z)/\phi}\right)dz + \frac{(H-z_1)(1-S_{wc})}{H} \tag{19}$$

***Fines-migration induced formation damage***: In the case of two-phase displacement, the capillary force exerted by the attached fine particles at the water-gas meniscus greatly exceeds the detaching drag force and the maximum attaching electrostatic force. As a result, the advanced water-gas meniscus detaches all detachable fines, i.e. the detrital fines attached to the rock surface by electrostatic forces. This process occurs in every capillary of the porous medium. Consequently, for each saturation value, the fines in small pores filled with water remain undetached, while all detachable fines in large pores filled with gas are detached [36]. Therefore, the concentration of retained particles in the porous media is a function of saturation. The attached fines concentration at saturation $s$ is given by:

$$\Delta\sigma_{cr} = \sigma_{cr}(0) - \sigma_{cr}(s), \quad \sigma_s(s) = \sigma_0 s^B \tag{20}$$

The fines attached by the meniscus migrate through the porous media and plug pore throats that are smaller than the particle size. Once plugged, the particle is bypassed by the displacing gas. As a result, the strained fines damage the gas phase but do not affect the displaced water. Consequently, the expression for phase velocities includes the formation damage factor for the gas phase only.

$$u_w = -\frac{k(z)k_{rw}(s)}{\mu_w}\frac{\partial p_w}{\partial r}, \quad u_g = -\frac{k(z)k_{rg}(s)}{\mu_g[1+\beta\Delta\sigma_{cr}(s)]}\frac{\partial p_g}{\partial r} \tag{21}$$

***Calculation of gas and water fluxes*:** Taking derivatives over *r* of both equations in equation (1) yields independence of pressure gradients in both phases of *z*. From equation (3) follows that capillary pressure gradient is also independent of *z*.

Water and gas fluxes are obtained by integrating the first and second equations of equation (21) over the reservoir thickness, respectively.

$$U_w = \int_0^H u_w dz = -\int_0^H \frac{k(z)k_{rw}(s)dz}{\mu_w}\frac{\partial p_w}{\partial r}, \quad U_g = \int_0^H u_g dz = -\int_0^H \frac{k(z)k_{rg}(s)dz}{\mu_g\left[1+\beta\Delta\sigma_{cr}(s)\right]}\frac{\partial p_g}{\partial r} \quad (22)$$

The overall rate is equal to the total of two rates:

$$Q = 2\pi r\left(U_w + U_g\right) = -2\pi r\int_0^H \frac{k(z)k_{rw}(s)dz}{\mu_w}\frac{\partial p_w}{\partial r} - 2\pi r\int_0^H \frac{k(z)k_{rg}(s)dz}{\mu_g\left[1+\beta\Delta\sigma_{cr}(s)\right]}\frac{\partial p_g}{\partial r}, \quad U = \frac{Q}{2\pi r} \quad (23)$$

Substituting the expression for pressure in gas from equation (22) and rearranging the terms yields:

$$Q = -2\pi r\left[\int_0^H \frac{k(z)k_{rw}(s)dz}{\mu_w} + \int_0^H \frac{k(z)k_{rg}(s)dz}{\mu_g\left[1+\beta\Delta\sigma_{cr}(s)\right]}\right]\frac{\partial p_w}{\partial r} - 2\pi r\int_0^H \frac{k(z)k_{rg}(s)dz}{\mu_g\left[1+\beta\Delta\sigma_{cr}(s)\right]}\frac{\partial P_c(s)}{\partial r} \quad (24)$$

Equation (24) takes the form:

$$U + \int_0^H \frac{k(z)k_{rg}(s)dz}{\mu_g\left[1+\beta\Delta\sigma_{cr}(s)\right]}\frac{\partial P_c(s)}{\partial r} = -\left[\int_0^H \frac{k(z)k_{rw}(s)dz}{\mu_w} + \int_0^H \frac{k(z)k_{rg}(s)dz}{\mu_g\left[1+\beta\Delta\sigma_{cr}(s)\right]}\right]\frac{\partial p_w}{\partial r} \quad (25)$$

Express pressure gradient in gas via that in water from equation (22) and calculate gas flux:

$$U_g = -\int_0^H \frac{k(z)k_{rg}(s)dz}{\mu_g\left[1+\beta\Delta\sigma_{cr}(s)\right]}\frac{\partial p_w}{\partial r} - \int_0^H \frac{k(z)k_{rg}(s)dz}{\mu_g\left[1+\beta\Delta\sigma_{cr}(s)\right]}\frac{\partial P_c(s)}{\partial r} \quad (26)$$

Expressing pressure gradient in water from equation (25) and substituting the results into equation (26) yields:

$$U_g = \int_0^H \frac{k(z)k_{rg}(s)dz}{\mu_g\left[1+\beta\Delta\sigma_{cr}(s)\right]} \frac{U + \int_0^H \frac{k(z)k_{rg}(s)dz}{\mu_g\left[1+\beta\Delta\sigma_{cr}(s)\right]} \frac{\partial P_c(s)}{\partial r}}{\left[\int_0^H \frac{k(z)k_{rw}(s)dz}{\mu_w} + \int_0^H \frac{k(z)k_{rg}(s)dz}{\mu_g\left[1+\beta\Delta\sigma_{cr}(s)\right]}\right]} - \int_0^H \frac{k(z)k_{rg}(s)dz}{\mu_g\left[1+\beta\Delta\sigma_{cr}(s)\right]} \frac{\partial P_c(s)}{\partial r} \qquad (27)$$

Now we introduce fractional flow function:

$$f = \frac{\int_0^H \frac{k(z)k_{rg}(s)dz}{\mu_g\left[1+\beta\Delta\sigma_{cr}(s)\right]}}{\left[\int_0^H \frac{k(z)k_{rw}(s)dz}{\mu_w} + \int_0^H \frac{k(z)k_{rg}(s)dz}{\mu_g\left[1+\beta\Delta\sigma_{cr}(s)\right]}\right]}, \qquad (28)$$

Expression for gas flux accounting for Eq. (28) becomes

$$U_g = Uf + \left\{ f\int_0^H \frac{k(z)k_{rg}(s)dz}{\mu_g\left[1+\beta\Delta\sigma_{cr}(s)\right]} - \int_0^H \frac{k(z)k_{rg}(s)dz}{\mu_g\left[1+\beta\Delta\sigma_{cr}(s)\right]} \right\} \frac{\partial P_c(s)}{\partial r} \qquad (29)$$

*Mass balance for gas*: Consider 2D flow in *r* and *z* directions. Conditions of impermeability are set at top and bottom of the reservoir, *z=H* and *z=0*, respectively. Integration of the continuity equation for both incompressible phases in z from z=0 to *z=H* accounting for incompressibility of the top and bottom boundaries results in conservation of the overall rate: *Q=Q(t)* [42]. Integration of the mass balance equation for gas in *z* from *z=0* to *z=H* accounting for incompressibility of the reservoir top and bottom yields:

$$r\phi H \frac{\partial S}{\partial t} + \frac{\partial}{\partial r}(rU_g) = 0, \quad S = \frac{1}{H}\int_0^H s(r,z,t)dz \qquad (30)$$

Substituting expression for gas phase from equation (29) yields:

$$r\phi H \frac{\partial S}{\partial t} + \frac{\partial}{\partial r}(rUf) = -\frac{\partial}{\partial r}\left\{ r\left[ f\int_0^H \frac{k(z)k_{rg}(s)dz}{\mu_g\left[1+\beta\Delta\sigma_{cr}(s)\right]} - \int_0^H \frac{k(z)k_{rg}(s)dz}{\mu_g\left[1+\beta\Delta\sigma_{cr}(s)\right]} \right] \frac{\partial P_c(s)}{\partial r} \right\} \qquad (31)$$

Further expansion of equation (31) and calculation of capillary pressure gradient from equation (11) yields:

$$\phi H \frac{\partial S}{\partial t} + \frac{Q}{2\pi r}\frac{\partial f}{\partial r} = \Delta\rho g \frac{\partial}{2r\partial r}\left\{4r^2\left[(1-f)\int_0^H \frac{k(z)k_{rg}(s)dz}{\mu_g\left[1+\beta\Delta\sigma_{cr}(s)\right]}\right]\frac{\partial z_0(r,t)}{2r\partial r}\right\} \quad (32)$$

Multiplying both sides of equation (32) by $R^2$ and dividing by $Q$ results in:

$$\frac{\pi\phi HR^2}{Q}\frac{\partial S}{\partial t} + \frac{1}{2r}\frac{\partial f}{\partial r}R^2 = \frac{\Delta\rho g\pi Hk_0}{Q\mu_g}R^2\frac{\partial}{2r\partial r}\left\{\frac{4r^2}{R^2}\left[(1-f)\int_0^H \frac{k(z)k_{rg}(s)dz}{k_0 H\left[1+\beta\Delta\sigma_{cr}(s)\right]}\right]\frac{\partial z_0(r,t)}{2r\partial r}\frac{R^2}{H}\right\} \quad (33)$$

Introduce the following dimensionless parameters:

$$x = \frac{r^2}{R^2},\ T = \frac{Qt}{\phi H\pi R^2},\ P = \frac{4\pi Hk_0}{\mu_w Q},\ Z = \frac{z}{H},\ K(Z) = \frac{k(z)}{k_0},\ h_c = \frac{H_c}{H},\ \varepsilon = \frac{4\pi\Delta\rho gHk_0}{\mu_g Q},\ P_{cD} = \frac{\sqrt{k_0/\phi}}{\sigma}P_c \quad (34)$$

Using the dimensionless coordinates and functions transforms equation (33) to the following form:

$$\frac{\partial S(Z_0)}{\partial T} + \frac{\partial f(Z_0)}{\partial x} = \varepsilon\frac{\partial}{\partial x}\left\{x\left[(1-f)\int_0^1 \frac{K(Z)k_{rg}(s)dZ}{\left[1+\beta\Delta\sigma_{cr}(s)\right]}\right]\frac{\partial Z_0(x,t)}{\partial x}\right\} \quad (35)$$

In zones 0 and 01, average saturation $S$ and fractional flow $f$ are expressed via the advanced front $Z_0$ by equations (17,18) and (28), respectively. Depth saturation distribution is also expressed via $Z_0$ by equation (12), so the integral in right hand side is a function of $Z_0$. The above closes equation (35) with unknown $Z_0(x,T)$. In zones 01 and 1, the average saturation $S$ and fractional flow $f$ are expressed via the receded front $Z_1$ by equations (19) and (28), respectively. The unknown in transformed equation (31) is $Z_1(x,T)$.

*Large scale approximation*: Parameter $\varepsilon$ is dimensionless ratio between the hydrostatic pressure and the pressure drop across the reservoir with radius $R$, which is considered to be significantly smaller than one:

$$\varepsilon = \frac{4\pi\Delta\rho gH^2 k_0}{\mu_g Q} \ll 1 \quad (36)$$

This allows neglecting right hand side of equation (32) compared with second term in left hand side of equation (32):

$$\frac{\partial S(Z_0)}{\partial T} + \frac{\partial f(Z_0)}{\partial x} = 0 \tag{37}$$

This Buckley-Leverett equation describes displacement of water by gas in zones 0 and 01. The parametric dependency of $Z_0$ by equation (17,18) and (28) allows $f$ or traditional fractional flow $f=f(S)$. The parametric dependency of $Z_1$ by equations (19) and (28) allows for traditional fractional flow $f=f(S)$ in zones 01 and 1. The expressions for function $f(S)$ as expressed via $Z_0$ and $Z_1$ in zone 01 coincides.

*2.2.1.1 Calculation of average saturation*

Saturation ahead of the advanced front $s=0$, so the average saturation is equal to zero too (Figure 2a). In zone 0, the vertical saturation column consists of the zone $S=0$ and the transition capillary zone. So,

$$S_0(x,T,Z_0) = \int_{Z_0}^{1} s(Z)dZ = \int_{Z_0}^{1} j\left(J_d - \Delta\rho g(Z-Z_0)\sqrt{K(Z)/\phi}\right)dZ \tag{38}$$

Saturation behind the receded front is equal to $1-S_{wc}$. In zone 01, the vertical saturation column consists of the zone $S=0$, the transition capillary zone and zone with $S=1-S_{wc}$. So,

$$\begin{aligned}S_{01}(x,T,Z_0) &= \int_{Z_0}^{Z_0+h_c} S(Z)dZ + (1-h_c-Z_0)(1-S_{wc}) \\ &= \int_{Z_0}^{Z_0+h_c} j\left(J_d - \Delta\rho g(Z-Z_0)\sqrt{K(Z)/\phi}\right)dZ + (1-h_c-Z_0)(1-S_{wc})\end{aligned} \tag{39}$$

In zone 1, the vertical saturation column consists of the transition capillary zone and zone with $S=1-S_{wc}$. So,

$$\begin{aligned}S_1(x,t,Z_1) &= \int_0^{Z_1} S(Z)dZ + (1-Z_1)(1-S_{wc}) \\ &= \int_0^{Z_1} j\left(J_m - \Delta\rho g(Z-Z_1)\sqrt{K(Z)/\phi}\right)dZ + (1-Z_1)(1-S_{wc})\end{aligned} \tag{40}$$

So, average saturations over the reservoir thickness for the above three cases are given by the formulae (38)-(40).

*2.2.1.2 Calculation of fractional flow function*

Let us calculate fractional flow function for three cases. In the case 0, gas flow occurs only in transition zone, between the reservoir top and advanced front:

$$f_0(Z_0) = \frac{\int_{Z_0}^{1} \frac{K(Z)k_{rg}(s)dZ}{\mu_g[1+\beta\Delta\sigma_{cr}(s)]}}{\left[\frac{1}{\mu_w}\int_{0}^{Z_0} K(Z)dZ + \int_{Z_0}^{1} \frac{K(Z)k_{rg}(s)dZ}{\mu_g[1+\beta\Delta\sigma_{cr}(s)]}\right]} \quad (41)$$

where equation (12) given the $Z_0$-dependent depth saturation distribution.

Gas flow in case 01 occurs in the transition zone between advanced and receded fronts, and above the receded front until the reservoir top, so gas flux contains two components:

$$f_{01}(Z_0) = \frac{\int_{Z_0}^{Z_0+h_c} \frac{K(Z)k_{rg}(s)dZ}{\mu_g[1+\beta\Delta\sigma_{cr}(s)]} + \left(\frac{k_{rgwc}}{\mu_g[1+\beta\Delta\sigma_{cr}(1-S_{wc})]}\int_{Z_0+h_c}^{1} K(Z)dZ\right)}{\left[\frac{1}{\mu_w}\int_{0}^{Z_0} K(Z)dZ + \int_{Z_0}^{Z_0+h_c}\frac{k_{rw}(s)K(Z)dZ}{\mu_w} + \int_{Z_0}^{Z_0+h_c}\frac{K(Z)k_{rg}(s)dZ}{\mu_g[1+\beta\Delta\sigma_{cr}(s)]} + \left(\frac{k_{rgwc}}{\mu_g[1+\beta\Delta\sigma_{cr}(1-S_{wc})]}\int_{Z_0+h_c}^{1} K(Z)dZ\right)\right]} \quad (42)$$

Here depth gas saturation is described by either equation (12) as $s=s(x,Z,T,Z_0)$ or by equation (13) as $s=s(x,Z,T,Z_1)$.

In the case 1, gas flows in the transition zone from the bottom inside the transition zone, and between the receded front and the reservoir top.

$$f_1(Z_1) = \frac{\int_{0}^{Z_1} \frac{K(Z)k_{rg}(s)dZ}{\mu_g[1+\beta\Delta\sigma_{cr}(s)]} + \left(\frac{k_{rgwc}}{\mu_g[1+\beta\Delta\sigma_{cr}(1-S_{wc})]}\int_{Z_1}^{1} K(Z)dZ\right)}{\left[\int_{0}^{Z_1}\frac{K(Z)k_{rw}(s)dZ}{\mu_w} + \int_{0}^{Z_1}\frac{K(Z)k_{rg}(s)dZ}{\mu_g[1+\beta\Delta\sigma_{cr}(s)]} + \left(\frac{k_{rgwc}}{\mu_g[1+\beta\Delta\sigma_{cr}(1-S_{wc})]}\int_{Z_1}^{1} K(Z)dZ\right)\right]} \quad (43)$$

Here, depth gas saturation is described by equation (13) as $s=s(x,Z,T,Z_1)$.

### 2.2.2 Governing equation in thin reservoirs ($H<H_c$)

Similar to thick reservoirs, we consider three cases: 0, 01, and 1, as defined in equation (10). In Case 0, the point *x* is located between the advanced front at $z=H$ and $z=0$. In Case 01, *x* lies between the receded front at $z=H$ and the advanced front at $z=0$. In Case 1, *x* is positioned between the receded front at $z=H$ and $z=0$ (Figure 2b). The following sections present the average saturation and fractional flow formulae for the three cases mentioned above in thin reservoirs.

#### 2.2.2.1 Calculation of average saturation

Equations (12) and (14) represent the saturation distribution along the vertical axis as a function of the positions of the advanced and receded fronts, respectively. To calculate the average saturation, equation (16) is applied, similar to thick reservoirs. Implementing this equation for thin reservoirs and using the dimensionless parameters in equation (34) yields the following average saturations over the reservoir thickness for the three cases mentioned above:

$$S_0(x,T,Z_0) = \int_{Z_0}^{1} j\left( J_d - \Delta\rho g(Z-Z_0)\sqrt{K(Z)/\phi} \right) dZ \tag{44}$$

$$S_{01}(x,T) = \int_{0}^{1} j\left( J_d - \Delta\rho g(Z-Z_0)\sqrt{K(Z)/\phi} \right) dZ \tag{45}$$

$$S_1(x,t,Z_1) = \int_{0}^{z_1} j\left( J_m - \Delta\rho g(Z-Z_1)\sqrt{K(Z)/\phi} \right) dZ + (1-Z_1)(1-S_{wc}) \tag{46}$$

Thus, the average saturations over the reservoir thickness for the three cases in thin reservoirs are given by equations (44) – (46).

#### 2.2.2.2 Calculation of fractional flow function

Let us calculate the fractional flow function for the three cases in thin reservoirs. In Case 0, gas flow occurs only in the transition zone between the reservoir top and the advanced front. Thus, the fractional flow in Zone 0 can be calculated using the following formulae:

$$f_0(Z_0) = \frac{\int_{Z_0}^{1} \frac{K(Z) k_{rg}(s) dZ}{\mu_g [1 + \beta \Delta \sigma_{cr}(s)]}}{\left[ \frac{1}{\mu_w} \int_0^{Z_0} K(Z) dZ + \int_{Z_0}^{1} \frac{K(Z) k_{rg}(s) dZ}{\mu_g [1 + \beta \Delta \sigma_{cr}(s)]} \right]} \qquad (47)$$

where equation (12) gives the $Z_0$-dependent depth saturation distribution.

In Case 01, gas flow occurs in the transition zone across the entire height of the reservoir. Therefore, the gas flux can be calculated as follows:

$$f_{01}(Z) = \frac{\int_0^1 \frac{K(Z) k_{rg}(s) dZ}{\mu_g [1 + \beta \Delta \sigma_{cr}(s)]}}{\left[ \int_0^1 \frac{k_{rw}(s) K(Z) dZ}{\mu_w} + \int_0^1 \frac{K(Z) k_{rg}(s) dZ}{\mu_g [1 + \beta \Delta \sigma_{cr}(s)]} \right]} \qquad (48)$$

In this region, $Z_0$ is equal to zero, and the fractional flow is calculated over the entire height of the reservoir, making it independent of $Z_0$.

In Case 1, gas flow takes place within the transition zone and extends between the receded front and the reservoir top. The fractional flow for this case is calculated as follows:

$$f_1(Z_1) = \frac{\int_0^{Z_1} \frac{K(Z) k_{rg}(s) dZ}{\mu_g [1 + \beta \Delta \sigma_{cr}(s)]} + \frac{k_{rgwc}}{\mu_g [1 + \beta \Delta \sigma_{cr}(1 - S_{wc})]} \int_{Z_1}^{1} K(Z) dZ}{\left[ \int_0^{Z_1} \frac{K(Z) k_{rw}(s) dZ}{\mu_w} + \int_0^{Z_1} \frac{K(Z) k_{rg}(s) dZ}{\mu_g [1 + \beta \Delta \sigma_{cr}(s)]} + \frac{k_{rgwc}}{\mu_g [1 + \beta \Delta \sigma_{cr}(1 - S_{wc})]} \int_{Z_1}^{1} K(Z) dZ \right]} \qquad (49)$$

Here, depth gas saturation is described by equation (13) as $s=s(x,Z,T,Z_1)$.

## 2.3 Calculation of well impedance

### 2.3.1 Buckley–Leverett solution

The following Buckley–Leverett equation is considered:

$$\frac{\partial s}{\partial T} + \frac{\partial f(s)}{\partial x} = 0 \qquad (50)$$

Considering the following conditions:

$$T = 0: \quad s = 0$$

$$x = x_w = \left(\frac{r_w}{R}\right)^2 : \quad s = 1 - S_{wc} \tag{51}$$

the solution for this equation can be expressed as follows:

$$s(x,T) = \begin{cases} 1 - S_{wc}, & \frac{x_w}{T} < \frac{x}{T} < D_i \\ \frac{x}{T} = f'(s), & D_i < \frac{x}{T} < D_f \\ 0, & D_f < \frac{x}{T} < \infty \end{cases} \tag{52}$$

In which:

$$D_i = f'(1 - S_{wi}), \quad D_f = f'(s_f) = \frac{f(s_f)}{s_f} \tag{53}$$

During the displacement of water by $CO_2$, a mixture zone forms between the pure $CO_2$ front and the water front. The average saturation across the medium is determined at fixed points along $x$, followed by the calculation of the fractional flow. Using this approach, the fractional flow consists of a shock region and a rarefied region, as defined in equation (52). If there is no transition or mixture zone, the displacement of water by $CO_2$ is entirely characterised by a shock. And the solution for shock can be expressed as follows:

$$s(x,T) = \begin{cases} 1 - S_{wc}, & 0 < \frac{x}{T} < D_s \\ 0, & D_s < \frac{x}{T} < \infty \end{cases} \tag{54}$$

In which:

$$D_s = \frac{f(s_f) - f(1 - S_{wc})}{s_f - (1 - S_{wc})} \tag{55}$$

### 2.3.2 Equations for gas and water fluxes

Equation (22) represents the water and gas fluxes. The overall flux, as expressed in equation (23), is the sum of the integrals of both fluxes presented in equation (22) along the depth. By considering the same pressure drop for gas and water in a large-scale approximation and applying the dimensionless parameter in equation (34), the following dimensionless pressure can be obtained:

$$1 = -\Lambda(s)\frac{\partial P}{\partial x} \tag{56}$$

In which, $\Lambda(s)$ is the total mobility which is defined as follows:

$$\Lambda(s) = \left[\int_0^1 K(Z)k_{rw}(s)dZ + \frac{\mu_w}{\mu_g}\int_0^1 \frac{K(Z)k_{rg}(s)dZ}{\left[1+\beta\Delta\sigma_{cr}(s)\right]}\right] \tag{57}$$

### 2.3.3 Exact solution

This section presents the exact solution for well impedance during the injection of $CO_2$ into aquifers, considering two immiscible and incompressible phases. It also accounts for the low compressibility of water ahead of the advancing $CO_2$ front. Based on equation (50), the equation for the dimensionless pressure drop can be written as:

$$\Delta P = \int_{x_w}^{\infty}\left(-\frac{\partial P}{\partial x}\right)dx = \int_{x_w}^{\infty}\frac{1}{x\Lambda(s)}dx \tag{58}$$

The pressure drop is calculated for the case before gas breakthrough, resulting in three distinct regions: the gas zone with irreducible water saturation, the two-phase zone, and the water zone ahead of the $CO_2$ front, as follows:

$$\Delta P(T) = \underbrace{\int_{x_w}^{D_tT}\frac{1}{x\Lambda(1-S_{wc})}dx}_{Gas\,Zone} + \underbrace{\int_{D_tT}^{D_fT}\frac{1}{x\Lambda(s)}dx}_{Two-Phase\,Zone} + \underbrace{\int_{D_fT}^{\infty}\frac{dP}{dx}dx}_{Water\,Zone,\,Compressible\,Part} = I_1 + I_2 + I_3 \tag{59}$$

The calculations for each component of equation (59) are presented below.

- *Gas zone with irreducible water saturation* ($I_1$)

The saturation in this zone is $s=1-S_{wc}$. Using this saturation, the first part of the integral in equation (59) can be calculated as follows:

$$I_1 = \frac{1}{\left[\dfrac{\mu_w}{\mu_g}\dfrac{k_{rgwc}}{\left[1+\beta\Delta\sigma_{cr}\left(1-S_{wc}\right)\right]}\int_0^1 K(Z)dZ\right]} \ln\left(\frac{D_iT}{x_w}\right) \quad (60)$$

where, $D_i$ can be calculated using equation (53).

- *Two-phase region* ($I_2$)

Using the solution for the Buckley–Leverett equation presented in equation (52), the following self-similar parameter can be defined:

$$\xi = \frac{x}{T} = f'(s) \rightarrow \begin{cases} x = \xi T \rightarrow dx = Td\xi \\ d\xi = f''(s)ds \end{cases} \quad (61)$$

Applying equation (61), the expression for the second integral of equation (59) in the two-phase gas-water region can be derived as follows:

$$I_2 = \int_{1-S_{wc}}^{0} \frac{f''(s)}{f'(s)\Lambda(s)} ds \quad (62)$$

- *Water zone ahead of the water-gas front* ($I_3$)

In this region, slight compressibility for water is considered, and the pressure diffusivity equation is solved. The diffusivity equation in radial coordinates is as follows [43]:

$$\frac{1}{r}\frac{\partial}{\partial r}\left(r\frac{\partial p}{\partial r}\right) = \frac{\varphi\mu_w c_t}{k}\frac{\partial p}{\partial t} \quad (63)$$

Considering the layer-caked permeability and the dimensionless variables presented in equations (34), the dimensionless form of the pressure diffusivity equation can be obtained as follows:

$$\frac{\partial}{\partial x}(x\frac{\partial P}{\partial x}) = \frac{Q\mu_w c_t}{4\pi H k_o}\frac{\partial P}{\partial T} \tag{64}$$

The method of self-similar solution is used to solve equation (64). To express equation (64) in terms of the self-similar parameter, the derivative of the self-similar parameter, as provided in equation (61), has been calculated with respect to $x$ and $T$ as follows:

$$\xi = \frac{x}{T} \rightarrow \begin{cases} \frac{d\xi}{dx} = \frac{1}{T} \\ \frac{d\xi}{dT} = -\frac{x}{T^2} \end{cases} \tag{65}$$

The derivation of the dimensionless pressure in equation (64) with respect to dimensionless time and position as a function of the self-similar parameter can be obtained as follows:

$$\frac{\partial P}{\partial T} = \frac{\partial P}{\partial \xi}\frac{\partial \xi}{\partial T} = -\frac{x}{T^2}\frac{\partial P}{\partial \xi} \tag{66}$$

$$\frac{\partial P}{\partial x} = \frac{\partial P}{\partial \xi}\frac{\partial \xi}{\partial x} = \frac{1}{T}\frac{\partial P}{\partial \xi} \tag{67}$$

Substituting equations (66) and (67) into equation (64), the pressure diffusivity equation will change to the following equation as a function of the self-similar parameter:

$$\frac{\partial}{\partial \xi}(\xi\frac{\partial P}{\partial \xi}) = -\left(\frac{Q\mu_w c_t}{4\pi H k_o}\right)\xi\frac{\partial P}{\partial \xi} \tag{68}$$

The following boundary conditions were applied to solve this equation:

$$\begin{aligned} &\xi = \infty : P = P_{reservoir} \\ &\xi = D_f : Q = Q_{inj} \end{aligned} \tag{69}$$

The first boundary condition in equation (69) implies that the pressure at a location far from the injection well is equal to the reservoir pressure. The second boundary condition states that the rate at the advanced gas front is equal to the injection rate. This condition arises from the assumption that gas is considered an incompressible fluid.

Expanding equation (68) and solving it using the method of separation of variables results in the following ordinary differential equation for the dimensionless pressure as a function of the self-similar parameter:

$$\frac{\partial P}{\partial \xi} = \frac{c_1}{\xi} e^{\left(-\left(\frac{Q\mu_w c_t}{4\pi H k_o}\right)\xi\right)} \tag{70}$$

By integrating both sides of equation (70) with respect to the self-similar variable, the following equation for the dimensionless pressure ahead of the gas front can be obtained as follows:

$$P = c_1 Ei\left(-\left(\frac{Q\mu_w c_t}{4\pi H k_o}\right)\xi\right) + c_2 \tag{71}$$

By applying the first boundary condition (69) to the solution (71), the value of $c_2$ can be obtained as follows:

$$P = c_1 Ei\left(-\left(\frac{Q\mu_w c_t}{4\pi H k_o}\right)\xi\right) + P_{reservoir} \tag{72}$$

To obtain $c_1$, the second boundary condition (69) should be applied. The following dimensionless Darcy equation can be expressed at the injection well as:

$$\frac{\partial P}{\partial x} = -\frac{1}{\left[\frac{\mu_w}{\mu_g}\frac{k_{rgwc}}{\left[1+\beta\Delta\sigma_{cr}(1-S_{wc})\right]}\int_0^1 K(Z)dZ\right]}\frac{1}{x} \tag{73}$$

Expressing equation (73) in terms of self-similar parameter (65):

$$\frac{\partial P}{\partial \xi} = -\frac{1}{\left[\frac{\mu_w}{\mu_g}\frac{k_{rgwc}}{\left[1+\beta\Delta\sigma_{cr}\left(1-S_{wc}\right)\right]}\int_0^1 K(Z)dZ\right]}\frac{1}{\xi} \qquad (74)$$

By applying the second condition (69) and equating equations (70) and (74), $c_1$ can be calculated as follows:

$$c_1 = -\frac{1}{\left[\frac{\mu_w}{\mu_g}\frac{k_{rgwc}}{\left[1+\beta\Delta\sigma_{cr}\left(1-S_{wc}\right)\right]}\int_0^1 K(Z)dZ\right]}e^{\left(\left(\frac{Q\mu_w c_t}{4\pi Hk_o}\right)D_f\right)} \qquad (75)$$

And the final form of the pressure solution in the water zone can be obtained as follows:

$$P = -\frac{e^{\left(\left(\frac{Q\mu_w c_t}{4\pi Hk_o}\right)D_f\right)}}{\left[\frac{\mu_w}{\mu_g}\frac{k_{rgwc}}{\left[1+\beta\Delta\sigma_{cr}\left(1-S_{wc}\right)\right]}\int_0^1 K(Z)dZ\right]}Ei\left(-\left(\frac{Q\mu_w c_t}{4\pi Hk_o}\right)\xi\right)+P_{reservoir} \qquad (76)$$

Using equation (76), the third integral in equation (59) can be calculated as follows:

$$I_3 = \frac{e^{\left(\left(\frac{Q\mu_w c_t}{4\pi Hk_o}\right)D_f\right)}}{\left[\frac{\mu_w}{\mu_g}\frac{k_{rgwc}}{\left[1+\beta\Delta\sigma_{cr}\left(1-S_{wc}\right)\right]}\int_0^1 K(Z)dZ\right]}Ei\left(-\left(\frac{Q\mu_w c_t}{4\pi Hk_o}\right)D_f\right) \qquad (77)$$

Substituting equations (60), (62), and (77) into equation (59) results in the total pressure drop across the three regions as follows:

$$\Delta P(T) = \frac{1}{\left[\frac{\mu_w}{\mu_g}\frac{k_{rgwc}}{\left[1+\beta\Delta\sigma_{cr}(1-S_{wc})\right]}\int_0^1 K(Z)dZ\right]}\ln\left(\frac{D_iT}{x_w}\right) + \int_{1-S_{wc}}^0 \frac{f''(s)}{f'(s)\Lambda(s)}ds + \ldots$$

$$\ldots + \frac{e^{\left(\left(\frac{Q\mu_w c_t}{4\pi H k_o}\right)D_f\right)}}{\left[\frac{\mu_w}{\mu_g}\frac{k_{rgwc}}{\left[1+\beta\Delta\sigma_{cr}(1-S_{wc})\right]}\int_0^1 K(Z)dZ\right]} Ei\left(-\left(\frac{Q\mu_w c_t}{4\pi H k_o}\right)D_f\right) \tag{78}$$

Well impedance, which is the normalised reciprocal to injectivity index, can be expressed in terms of pressure drop as follows [44]:

$$J(T) = \frac{II(T=0)}{II(T)} = \frac{Q(T=0)}{\Delta P(T=0)}\frac{\Delta P(T)}{Q(T)} \tag{79}$$

Considering constant rate injection and the following initial dimensionless pressure drop:

$$\Delta P(0) = -\ln(x_w) \tag{80}$$

The well impedance is calculated by dividing equation (78) by equation (80) as follows:

$$J(T) = -\frac{1}{\ln(x_w)\left[\frac{\mu_w}{\mu_g}\frac{k_{rgwc}}{\left[1+\beta\Delta\sigma_{cr}(1-S_{wc})\right]}\int_0^1 K(Z)dZ\right]}\ln\left(\frac{D_iT}{x_w}\right) + \frac{1}{\ln(x_w)}\int_0^{1-S_{wc}}\frac{f''(s)}{f'(s)\Lambda(s)}ds - \ldots$$

$$\ldots - \frac{e^{\left(\left(\frac{Q\mu_w c_t}{4\pi H k_o}\right)D_f\right)}}{\ln(x_w)\left[\frac{\mu_w}{\mu_g}\frac{k_{rgwc}}{\left[1+\beta\Delta\sigma_{cr}(1-S_{wc})\right]}\int_0^1 K(Z)dZ\right]} Ei\left(-\left(\frac{Q\mu_w c_t}{4\pi H k_o}\right)D_f\right) \tag{81}$$

## 2.4 Sweep calculations

Now, the average water saturation is calculated using the following expression and the exact solution (52).

$$\bar{s}(T) = \int_{x_w}^1 s(x,T)dx \tag{82}$$

To calculate the average water saturation as a function of T, defined by equation (82), equation (52) is integrated over the triangular domain $\Gamma: (x_w, 0) \to (x_w, T) \to (1, T) \to (x_w, 0)$ in Figure 3.

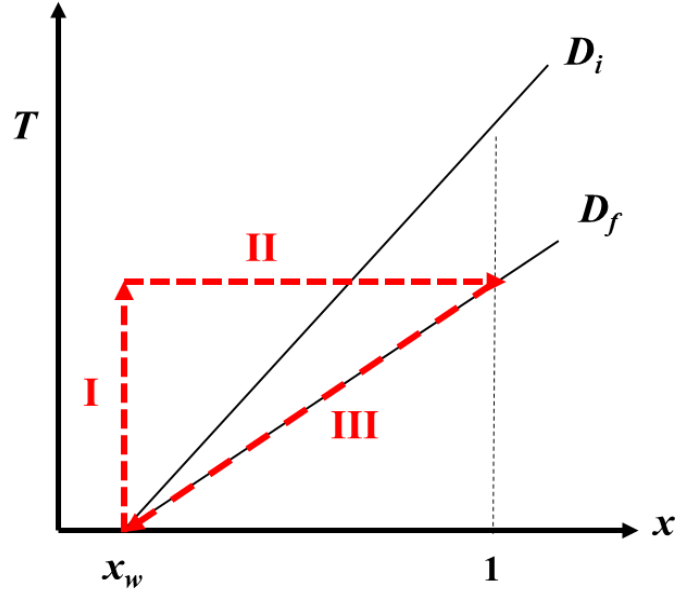

Figure 3. Structure of flow zone at plane (*x*, *T*).

According to Green's theorem, this integral is equivalent to a contour integral over the domain boundary $\partial \Gamma$ of the flux $f dt_D - S dx_D$, as follows:

$$\iint \left[ \frac{\partial s}{\partial T} + \frac{\partial f}{\partial x} \right] dT dx = \oint f dT - s dx = 0 \tag{83}$$

First, consider the period after the arrival of the advanced gas front at the drainage radius and before the arrival of the receded gas front, $1/D_f < T < 1/D_i$. The contour integral over the triangular domain can be decomposed into three integrals, as follows:

$$\oint f dT - s dx = \int_{(x_w,0)}^{(x_w,T)} + \int_{(x_w,T)}^{(1,T)} + \int_{(1,T)}^{(x_w,0)} = 0 \tag{84}$$

Each integral can now be calculated as follows. the first integral, where the integration is performed along $T$, the term involving $dx$ is zero. Consequently, the integrand $f(1-S_{wc})$ is equal to $T$.

$$\int_{(x_w,0)}^{(x_w,T)} f(s(0))dT - s(0)dx = \underbrace{\int_0^T f(s)dT}_{@ x=x_w, s=1-S_{wc}, f=1} - \int_{x_w}^{x_w} s\, dx = T \tag{85}$$

In the second integral, where the integration is carried out along $x$, the term involving $dT$ is equal to zero. As stated in equation (82), the integral with respect to $x$ provides the average water saturation.

$$\int_{(x_w,T)}^{(1,T)} f(s(\tfrac{x}{T}))dT - s(\tfrac{x}{T})dx = \underbrace{\int_T^T f(s(\tfrac{x}{T}))dT}_{0} - \int_{x_w}^1 s(\tfrac{x}{t})dx = -\bar{s}(T) \tag{86}$$

In the third integral, $s$ and $f$ remain constant along the hypotenuse of the triangular domain $\Gamma$.

$$\int_{(1,T)}^{(x_w,0)} f(s(\tfrac{1}{T}))dT - s(\tfrac{1}{T})dx = \int_T^0 f(s(\tfrac{1}{T}))dT - \int_1^{x_w} s(\tfrac{1}{T})dx = (1-x_w)s(\tfrac{1}{T}) - f(s(\tfrac{1}{T}))T \tag{87}$$

For times before the arrival of the advanced gas front, $T < 1/D_f$, the first integral, given by equation (85), remains equal to $T$, while the second integral, as expressed in equation (86), equals the negative of the average saturation. Along the hypotenuse of the triangle, where $s=f=0$, the third integral evaluates to zero in the $(x,T)$ plane.

For times after the arrival of the receded gas front, $T > 1/D_i$, the first integral in equation (85) remains equal to $T$, while the second integral in equation (86) is equal to the negative of the average saturation. Along the hypotenuse of the triangle, where $s=1-S_{wc}$ and $f=1$, the third integral equals $(1-x_w)(1-S_{wc})$ in the $(x,T)$ plane. The final expression for the average gas saturation at the three moments mentioned above is given by equation (88).

$$\bar{s}(T) = \begin{cases} T & 0 < T < \dfrac{1}{D_f} \\ (1-x_w)s(\dfrac{1}{T}) + \left(1 - f(s(\dfrac{1}{T}))\right)T & \dfrac{1}{D_f} < T < \dfrac{1}{D_i} \\ (1-x_w)(1-S_{wc}) & \dfrac{1}{D_f} < T < \infty \end{cases} \quad (88)$$

Equation (88) calculates the average gas saturation. The sweep is defined as the average gas saturation at $T=1$, which represents the area of the reservoir swept by the injected gas at $T=1$.

## 3 Practical calculations

Section 2 presents the governing equations, along with the general solutions obtained for depth saturation distribution, well impedance, and sweep efficiency. In this section, the specific expressions for capillary pressure, relative permeability, and the permeability profile are considered, as applied in the practical calculations.

### 3.1 Capillary pressure

In this section the normalised gas saturation is defined as $S = \dfrac{s}{1 - S_{wc} + \varepsilon}$ to ensure that saturation remains between 0 and 1. The parameter $\varepsilon$ is a small value approaching zero. It is introduced into the normalised saturation to resolve the singularity that arises when $s=1-S_{wc}$, which occurs during the calculation of maximum capillary pressure using the Brooks-Corey model, as depicted in Figure 1.

Using the dimensionless parameters (34), the Brooks-Corey model can be expressed as a function of normalised saturation as follows [45]:

$$P_{cD} = P_{d_D} (1-S)^{-\frac{1}{\lambda}} \quad (89)$$

Using equation (89), the capillary pressure at $s=0$ (which corresponds to $S=0$) represents the threshold pressure ($P_{d_D}$) at which gas begins to enter the largest pore of the reservoir. The maximum capillary pressure ($P_{m_D}$), occurring at $s=1-S_{wc}$ (which is $S \approx 1$), represents the region where gas has displaced water. Actually, by applying $\varepsilon$ mentioned above, equation (89) will be equal to $P_{m_D}$, not an infinite value. Both pressures are illustrated in Figure 1.

Applying equation (3) along with the dimensionless parameters (34), the following expression can be derived for the dimensionless capillary pressure and depth.

$$dP_{cD}(S) = -\gamma dZ, \quad \gamma = \frac{\Delta \rho g \sqrt{k_0/\phi} H}{\sigma} \tag{90}$$

Calculating the derivative of Equation (89) with respect to $S$ and substituting into Equation (90) will result in the relationship between dimensionless height and normalised saturation in the two-phase zone, as follows:

$$dZ = -\frac{P_{d_D}}{\gamma \lambda}(1-S)^{-\frac{1}{\lambda}-1} dS \tag{91}$$

3.2 Relative permeability

For relative permeabilities, Corey's formulae (92) and (93) are applied for water and gas, respectively [46].

$$k_{rw} = K_{rwgr}(1-S)^{3+\frac{2}{\lambda}} \tag{92}$$

$$k_{rg} = K_{rgwc}\left[(S)^2 \left(1-(1-S)^{1+\frac{2}{\lambda}}\right)\right] \tag{93}$$

## 3.3 Permeability profile along the depth

A linear permeability distribution is assumed for the permeability profile along the height of the reservoir, with permeability decreasing with depth. Based on this assumption, the permeability profile can be expressed as follows:

$$k(z) = (k_{min} - k_{max})\left(\frac{z}{H}\right) + k_{max} \rightarrow k(Z) = (k_{min} - k_{max})Z + k_{max} \tag{94}$$

Applying the dimensionless parameters (34), the following expression can be derived for the dimensionless permeability as a function of dimensionless depth:

$$K(Z) = \frac{2}{k_{min} + k_{max}} \left[(k_{min} - k_{max})Z + k_{max}\right] \tag{95}$$

For calculation of fractional flow, it is necessary to express the permeability profile as a function of saturation and the positions of the advanced ($Z_0$) and receded ($Z_1$) fronts. By using the constants obtained in Equation (8) and applying them to Equation (5) along with the dimensionless parameters (34), the following formulae can be derived to express $Z$ as a function of $Z_0$ and $Z_1$:

$$Z = \frac{P_{d_D} - P_{cD}}{\gamma} + Z_0 \tag{96}$$

$$Z = \frac{P_{m_D} - P_{cD}}{\gamma} + Z_1 \tag{97}$$

By substituting Equations (96) and (97) into Equation (95) and applying the definition of capillary pressure from Equation (89), Equations (98) and (99) are derived to express the permeability profile versus depth as a function of normalised saturation and the positions of the advanced and receded fronts, respectively.

$$K(S) = \frac{2}{k_{min} + k_{max}} \left[ (k_{min} - k_{max}) P_{d_D} \left( \frac{1 - (1-S)^{-\frac{1}{\lambda}}}{\gamma} + Z_0 \right) + k_{max} \right] \quad (98)$$

$$K(S) = \frac{2}{k_{min} + k_{max}} \left[ (k_{min} - k_{max}) \left( \frac{P_{m_D} - P_{d_D}(1-S)^{-\frac{1}{\lambda}}}{\gamma} + Z_1 \right) + k_{max} \right] \quad (99)$$

where equations (12) and (14) give the $Z_0$ and $Z_1$ dependent depth saturation distribution.

Using the above-mentioned equations for capillary pressure, relative permeability, and the permeability profile versus depth, the following section calculates the average saturation and fractional flow equations for different regions of thick and thin reservoirs.

### 3.4 Calculations for thick reservoirs

Figure 2a shows a schematic of thick reservoirs and the different regions based on the positions of the advanced and receded fronts, as described in equation (10). The following sections present the calculations for average saturations and fractional flow for thick reservoirs.

### 3.4.1 Calculation of depth saturation distribution

The average gas saturation in Zone 0, in which the vertical saturation column consists of the zone $S=0$ and the transition capillary zone, can be calculated as using equation (38). the pressure at the front of $Z_0$ corresponds to $P_{d_D}$, which is the threshold capillary pressure for entering the region with saturation $S=0$. Consequently, the capillary pressure at the top of the reservoir is determined by the pressure at the front $Z_0$, along with the hydrostatic pressure above the front, up to the top of the reservoir. Let the normalised saturation at the top of the reservoir be denoted as $S^*$. By substituting this saturation into equation (89) for the capillary pressure, the following expression can be obtained:

$$P_{d_D} + \gamma Z_0 = P_{d_D}\left(1-S^*\right)^{-\frac{1}{\lambda}} \rightarrow S^* = 1 - \left(\frac{P_{d_D} + \gamma Z_0}{P_{d_D}}\right)^{-\lambda} \tag{100}$$

$S^*$ represents the saturation of the gas at the top of the reservoir in Zone 0, based on the position of the advanced front. Normalising both sides of equation (38) by $1-S_{wc}$, and substituting $dZ$ from equation (91) into this equation, results in the following expression for average gas saturation in Zone 0.

$$S_0(x,T,Z_0) = -\frac{P_{d_D}}{\gamma\lambda}\int_0^{S^*} S(1-S)^{-\frac{1}{\lambda}-1} dS = \frac{P_{d_D}}{\gamma(\frac{1}{\lambda}-1)}\left(1+(1-S^*)^{-\frac{1}{\lambda}}\left(\frac{S^*}{\lambda}-1\right)\right) \tag{101}$$

The dependency of $S^*$ on $Z_0$ is shown in equation (100).

Let us calculate the average gas saturation for Zone 01, where $x$ is situated between the position of the receded front at $Z=1$ and the advanced front at $Z=0$. The saturation at $Z_0$ is equal to zero, and the saturation at $Z_0+h_C$ is equal to one. Using the same procedure applied for Zone 0, by normalising both sides of Equation (39) by $1-S_{wc}$ and substituting $dZ$ from equation (91) into equation (39), the following expression for average gas saturation in Zone 01 is obtained.

$$S_{01}(x,T,Z_0) = -\frac{P_{d_D}}{\gamma\lambda}\int_0^1 S(1-S)^{-\frac{1}{\lambda}-1} dS + (1-h_c-Z_0) = (1-h_c-Z_0) - \frac{1}{\gamma}\left(P_{m_D} + \frac{P_{d_D}}{(\frac{1}{\lambda}-1)}\right) \tag{102}$$

To calculate the average saturation in Zone 1, equation (40) is applied. The saturation at $Z_1$ is equal to $S=1$, and, similar to the approach used for Zone 0, in this zone, we need to calculate the saturation at the bottom of the reservoir in this region. This corresponds to the capillary pressure at the front $Z_1$, which is $P_{m_D}$ minus the hydrostatic head below the front $Z_1$. Let the normalised saturation at the bottom of the reservoir be denoted as $S_*$. By substituting this saturation into equation (89) for the capillary pressure, the following expression is obtained:

$$P_{m_D} - \gamma(1-Z_1) = P_{d_D}(1-S_*)^{-\frac{1}{\lambda}} \rightarrow S_* = 1 - \left(\frac{P_{m_D} - \gamma(1-Z_1)}{P_{d_D}}\right)^{-\lambda} \tag{103}$$

Normalising both sides of equation (40) by 1-$S_{wc}$, and substituting $dZ$ from equation (91) into this equation, results in the following expression for average gas saturation in Zone 1.

$$S_1(x,t,Z_1) = -\frac{P_{d_D}}{\gamma\lambda}\int_{S_*}^{1} S(1-S)^{-\frac{1}{\lambda}-1} dS + (1-Z_1) = (1-Z_1) - \frac{1}{\gamma}\left(P_{m_D} - \frac{P_{d_D}\left((1-S_*)^{-\frac{1}{\lambda}}\left(\frac{1}{\lambda}S_* - 1\right)\right)}{(\frac{1}{\lambda}-1)}\right) \tag{104}$$

Equations (101), (102), and (104) calculate the average gas saturation along the height for the three zones 01, 0, and 1 in thick reservoirs, respectively.

3.4.2 Calculation of fractional flow function

Let us begin by calculating the fractional flow for Zone 0. To do this, equation (41) is applied. By substituting equations (20), (91), (93), and (98) into equation (41), the following expression is obtained to calculate the fractional flow in Zone 0.

$$f_0(Z_0) = \frac{A}{A+B} \tag{105}$$

Where,

$$A = \int_{0}^{S^*} \frac{\left[(k_{min} - k_{max})P_{d_D}\left(\frac{1-(1-S)^{-\frac{1}{\lambda}}}{\gamma} + Z_0\right) + k_{max}\right]\left[(S)^2\left(1-(1-S)^{1+\frac{2}{\lambda}}\right)\left((1-S)^{-\frac{1}{\lambda}-1}\right)\right] dS}{\left[1+\beta\sigma_0 S^B\right]} \tag{106}$$

$$B = \frac{1}{K_{rgwc}}\frac{\mu_g}{\mu_w}\int_{0}^{Z_0}\left[(k_{min} - k_{max})P_{d_D}\left(\frac{1-(1-S)^{-\frac{1}{\lambda}}}{\gamma} + Z_0\right) + k_{max}\right](1-S)^{-\frac{1}{\lambda}-1} dS \tag{107}$$

Equation (100) shows the dependency of $S^*$ on $Z_0$.

To calculate the fractional flow in Zone 01, equation (42) is used. By applying the same substitutions as in Zone 0, the following expression is obtained for the calculation of fractional flow in Zone 01.

$$f_{01}(Z_0) = \frac{C}{C+D} \tag{108}$$

Where,

$$C = -\frac{P_{d_D}}{\gamma\lambda}\frac{K_{rgwc}}{\mu_g}\int_0^1 \frac{\left((k_{min}-k_{max})P_{d_D}\left(\frac{1-(1-S)^{-\frac{1}{\lambda}}}{\gamma}+Z_0\right)+k_{max}\right)\left[(S)^2\left(1-(1-S)^{1+\frac{2}{\lambda}}\right)(1-S)^{-\frac{1}{\lambda}-1}\right]}{\left[1+\beta\sigma_0 S^B\right]}dS + \ldots$$
$$\ldots + \left[\frac{k_{rgwc}}{\mu_g[1+\beta\sigma_0]}\left(\left(\frac{k_{min}+k_{max}}{2}\right)-(k_{min}-k_{max})\frac{(Z_0+h_c)^2}{2}-k_{max}(Z_0+h_c)\right)\right] \tag{109}$$

$$D = \frac{1}{\mu_w}\left((k_{min}-k_{max})\frac{Z_0^2}{2}+k_{max}Z_0\right) - \ldots$$
$$\ldots - \frac{P_{d_D}}{\gamma\lambda}\frac{K_{rwgr}}{\mu_w}\int_0^1\left[\left((1-S)^{3+\frac{2}{\lambda}}\right)\left((k_{min}-k_{max})P_{d_D}\left(\frac{1-(1-S)^{-\frac{1}{\lambda}}}{\gamma}+Z_0\right)+k_{max}\right)\left((1-S)^{-\frac{1}{\lambda}-1}\right)\right]dS \tag{110}$$

Applying the same procedure used above, the fractional flow in Zone 1 is obtained as follows:

$$f_1(Z_1) = \frac{E}{E+F} \tag{111}$$

Where,

$$E = \frac{K_{rgwc}}{\mu_g}\int_{S_*}^1 \frac{\left((k_{min}-k_{max})\left(\frac{P_{m_D}-P_{d_D}(1-S)^{-\frac{1}{\lambda}}}{\gamma}+Z_1\right)+k_{max}\right)\left[(S)^2\left(1-(1-S)^{1+\frac{2}{\lambda}}\right)\left((1-S)^{-\frac{1}{\lambda}-1}\right)\right]dS}{\left[1+\beta\sigma_0 S^B\right]} + \ldots$$
$$\ldots + \left(\frac{k_{rgwc}}{\mu_g[1+\beta\sigma_0]}\left((\frac{k_{min}+k_{max}}{2})-(k_{min}-k_{max})\frac{Z_1^2}{2}-k_{max}Z_1\right)\right) \tag{112}$$

$$F = \frac{K_{rwgr}}{\mu_w}\int_0^{Z_1}\left((k_{min}-k_{max})\left(\frac{P_{m_D}-P_{d_D}(1-S)^{-\frac{1}{\lambda}}}{\gamma}+Z_1\right)+k_{max}\right)\left((1-S)^{2+\frac{1}{\lambda}}\right)dS \tag{113}$$

Equation (100) shows the dependency of $S_*$ on $Z_1$.

Equations (105), (108), and (111) calculate the average gas saturation along the height for the three zones 01, 0, and 1 in thick reservoirs, respectively.

## 3.5 Calculations for thin reservoirs

Figure 2b illustrates a schematic of thick reservoirs, highlighting the different regions defined by the positions of the advanced and receded fronts, as described in equation (10). The subsequent sections provide the calculations for average saturations and fractional flow in thick reservoirs.

### 3.5.1 Calculation of depth saturation distribution

The depth saturation distribution calculations for Zones 0 and 1 in thin reservoirs are identical to those for thick reservoirs, with the corresponding equations provided in equations (101) and (104). Therefore, these calculations are not repeated here. The only difference between thick and thin reservoirs lies in Zone 01, for which the depth-dependent saturation calculation is presented below.

In Zone 01, two-phase flow occurs throughout the entire height of the reservoir. Similar to the approach used for this zone in thick reservoirs, where the saturation at the top of the reservoir was determined, the same is required here. Referring to Figure 2b, the capillary pressure at the top of the reservoir consists of the pressure at the front along with the hydrostatic pressure exerted above it, up to the reservoir top. In thin reservoirs, Zone 01 begins when $Z_0$ reaches the bottom of the reservoir. Consequently, the capillary pressure at the top of the reservoir is calculated as the capillary pressure at $Z_0$, denoted as $P_{d_D}$, plus the hydrostatic pressure exerted from the entire reservoir height. Let the normalised saturation at the top of the reservoir in Zone 01 for thin reservoirs, above the tail of the advanced front, be denoted as $S^{**}$. By substituting this saturation into equation (89) for capillary pressure, the following expression is obtained:

$$P_{d_D} + \gamma = P_{d_D}\left(1-S^*\right)^{-\frac{1}{\lambda}} \rightarrow S^{**} = 1 - \left(\frac{P_{d_D}+\gamma}{P_{d_D}}\right)^{-\lambda} \tag{114}$$

Normalising both sides of equation (45) by 1-$S_{wc}$, and substituting $dZ$ from equation (91) into this equation, results in the following expression for average gas saturation in Zone 01.

$$S_{01}(x,T) = -\frac{P_{d_D}}{\gamma\lambda}\int_0^{S^{**}} S(1-S)^{-\frac{1}{\lambda}-1}dS = -\frac{P_{d_D}}{\gamma(\frac{1}{\lambda}-1)}\left[\left(1-S^{**}\right)^{-\frac{1}{\lambda}}(\frac{S^{**}}{\lambda}-1)+1\right] \tag{115}$$

It should be noted that the saturation in this region can be obtained either through the tail of the advanced front, as shown in Equation (115), or through the head of the receded front, where the same methodology can be applied.

### 3.5.2 Calculation of fractional flow function

Similar to the calculation for saturation, the fractional flow in both regions 0 and 1 for thick and thin reservoirs is the same, as already presented in equations (105) and (111), respectively. So, here we only calculate the fractional flow in Zone 01. By substituting equations (89), (91), (92), (93), and (98) into equation (48), the following formula will be obtained for the fractional flow in Zone 01 of thin reservoirs.

$$f_{01}(Z) = \frac{L}{L+M} \tag{116}$$

Where,

$$L = \frac{K_{rgwc}}{\mu_g}\int_0^{S^{**}} \frac{\left((k_{min}-k_{max})P_{d_D}\left[\frac{1-(1-S)^{-\frac{1}{\lambda}}}{\gamma}\right]+k_{max}\right)\left[(S)^2\left(1-(1-S)^{1+\frac{2}{\lambda}}\right)\left((1-S)^{-\frac{1}{\lambda}-1}\right)\right]dS}{\left[1+\beta\sigma_0 S^B\right]} \tag{117}$$

$$M = \frac{K_{rwgr}}{\mu_w} \int_0^{S^{**}} \left( (k_{min} - k_{max}) P_{d_D} \left( \frac{1-(1-S)^{-\frac{1}{\lambda}}}{\gamma} \right) + k_{max} \right) \left( (1-S)^{\frac{1}{\lambda}+2} \right) dS \qquad (118)$$

which $S^{**}$ is presented in equation (114).

## 4 Analytical modelling results and sensitivity study

This section presents a sensitivity analysis of the key parameters incorporated into the developed model. Figure 4 illustrates a typical gas fractional flow with a shock front, obtained by the model developed in this study.

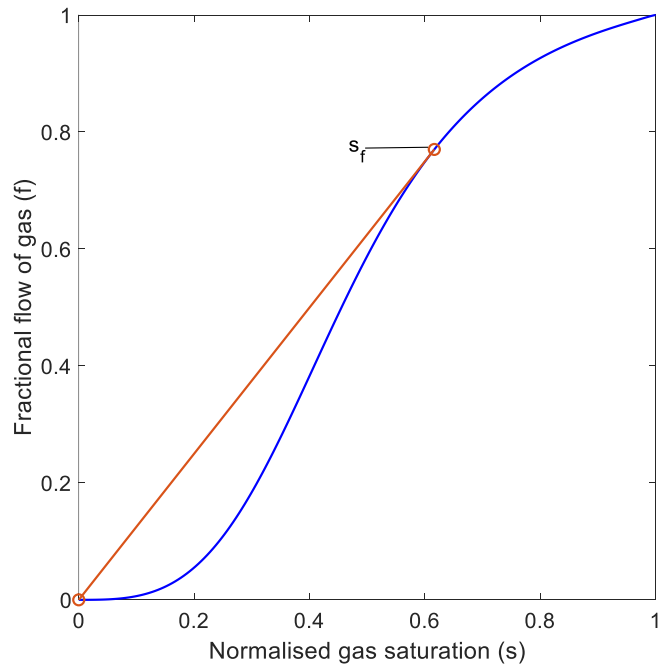

Figure 4. Fractional flow curve of gas with shock front.

Figure 5 illustrates the impact of the gas-water viscosity ratio ($\mu$) on the fractional flow curves. As shown, an increase in the gas-water viscosity ratio results in a decrease in gas fractional flow. This is due to the reduced mobility of the gas phase within the porous medium as gas viscosity increases. Higher gas viscosity increases flow resistance, making it harder for the gas to displace the liquid phase and reducing the relative permeability to gas. As a result, the gas fractional flow decreases.

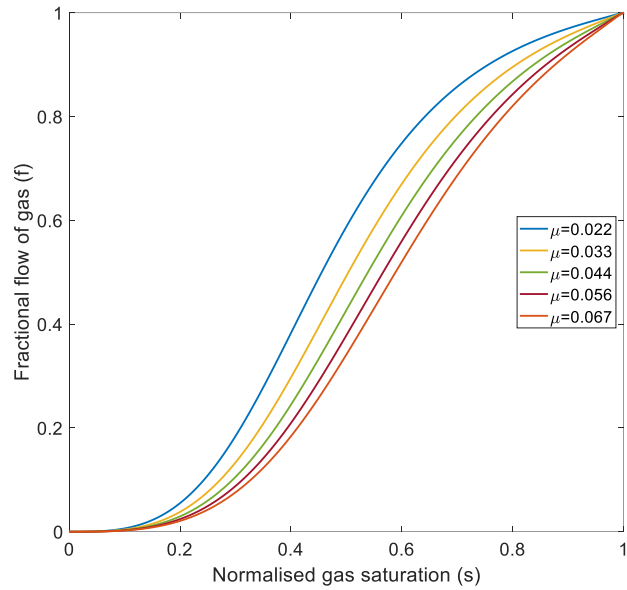

Figure 5. Fractional flow curve for varying gas-water viscosity ratio.

Figure 6 illustrates the impact of the gas-water viscosity ratio on the gas saturation profile. As observed, an increase in the gas-water viscosity ratio results in a higher gas saturation. This occurs because higher gas viscosity stabilises the gas front, reducing the occurrence of gas fingering through the water phase. A more stable gas front enhances the displacement of water by the gas, leading to more efficient sweep and consequently higher gas saturation within the reservoir.

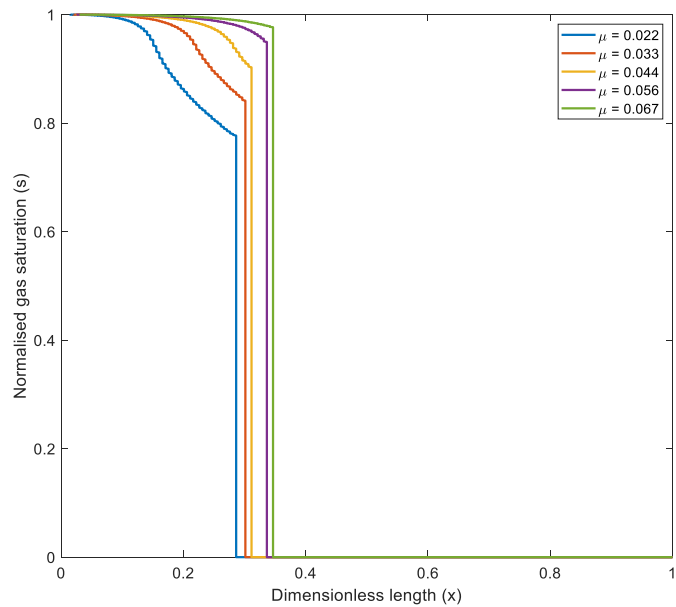

Figure 6. Gas saturation profile at moment T=0.5 for varying gas-water viscosity ratio.

Figure 7 illustrates the impact of formation damage caused by fines migration on gas fractional flow. Figure 7(a) presents the relationship between the formation damage coefficient ($\beta$) and gas fractional flow. As observed, an increase in $\beta$ leads to a reduction in gas fractional flow. This trend can be attributed to the assumption that fines migration occurs exclusively behind the gas front. Consequently, a higher $\beta$ value corresponds to greater formation damage within the gas zone, resulting in a decreased gas fractional flow. A similar trend is observed for $\sigma_0$, as depicted in Figure 7(b). Figure 7 also shows that the effect of $\beta$ on fractional flow is more pronounced than that of $\sigma_0$.

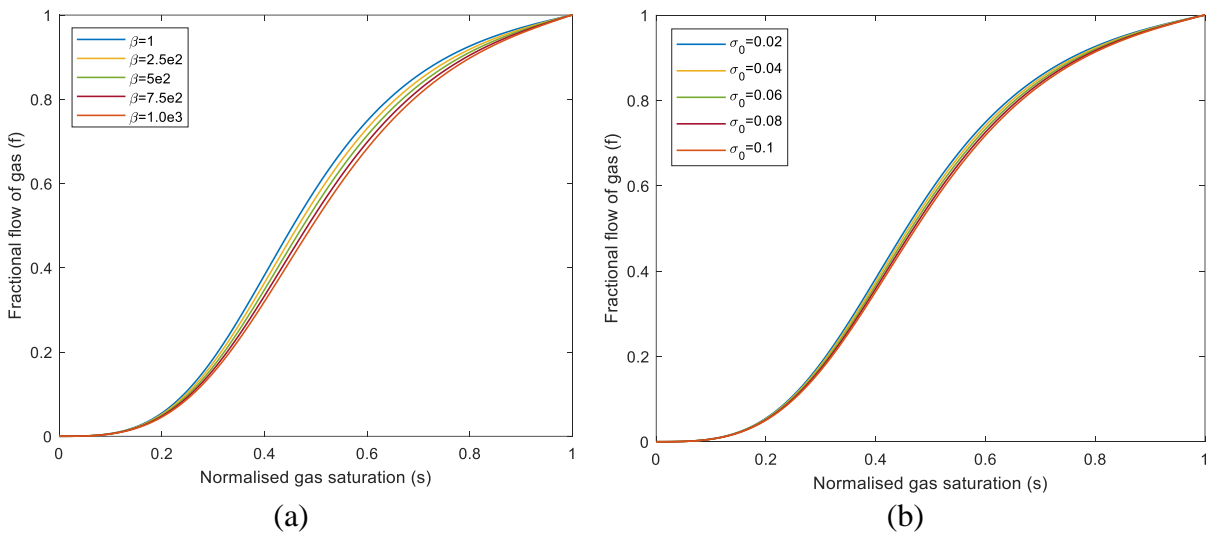

Figure 7. Fractional flow curve for varying (a) $\beta$ and (b) $\sigma_0$.

Figure 8 illustrates the impact of fines migration on the gas saturation profile. Figure 8(a) demonstrates that increasing the $\beta$ value results in higher gas saturation. A similar trend is observed with variations in $\sigma_0$, as shown in Figure 8(b). This phenomenon occurs because the migrated fines, within the gas zone, are strained in narrow pore throats. As a result, the fines become trapped, hindering the flow of the injected gas and causing formation damage, which in turn leads to a decline in gas phase permeability (please see the Figure 7). This reduction in gas phase permeability redirects the gas flow to unswept areas of the reservoir, thereby improving sweep efficiency and increasing the overall gas saturation.

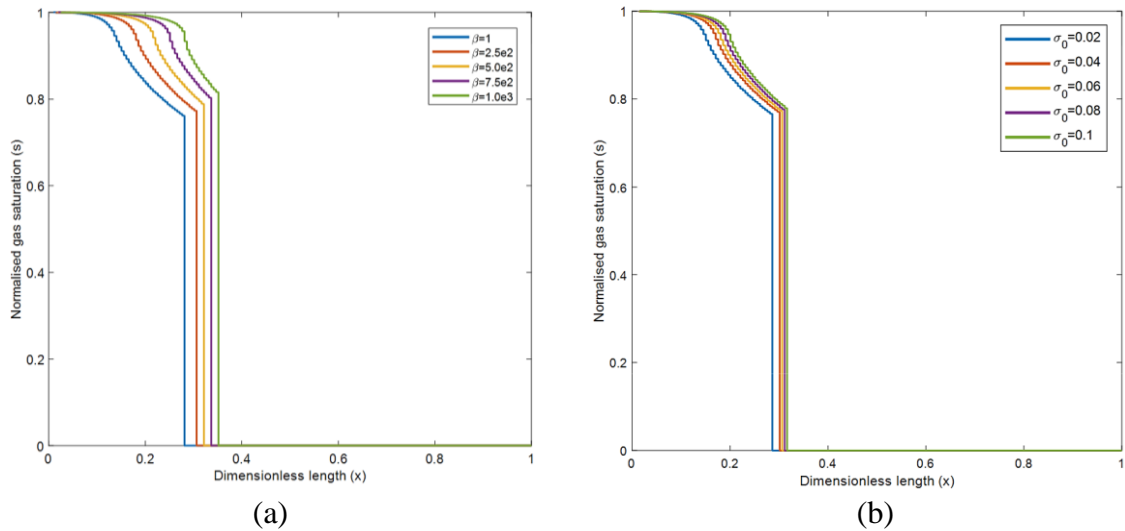

(a)                                                   (b)

Figure 8. Gas saturation profile at moment $t=0.5$ PVI for varying (a) $\beta$ and (b) $\sigma_0$.

Figure 9 illustrates the impact of formation damage caused by fines migration on sweep efficiency. Sweep efficiency is defined as the average gas saturation after one pore volume of gas injection, as calculated by equation (125). The figure clearly demonstrates that an increase in both $\beta$ and $\sigma_0$ leads to an improvement in sweep efficiency. These findings are consistent with the results presented in Figure 10.

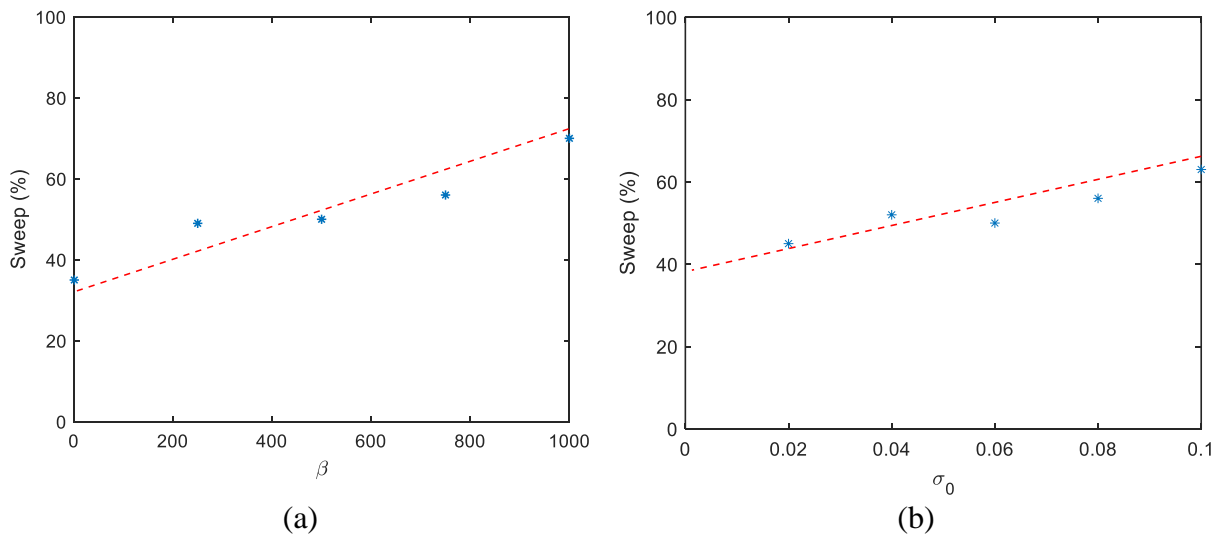

(a)                                                   (b)

Figure 9. Sweep efficiency for varying (a) $\beta$ and (b) $\sigma_0$.

Figure 10 illustrates the effect of fines migration on well injectivity. The figure shows that increasing $\beta$ and $\sigma_0$ increases the well impedance, which in turn reduces the well injectivity. These results indicate that while fines migration enhances sweep efficiency and increases $CO_2$ storage capacity, it simultaneously decreases well injectivity, which should be considered for successful underground $CO_2$ storage.

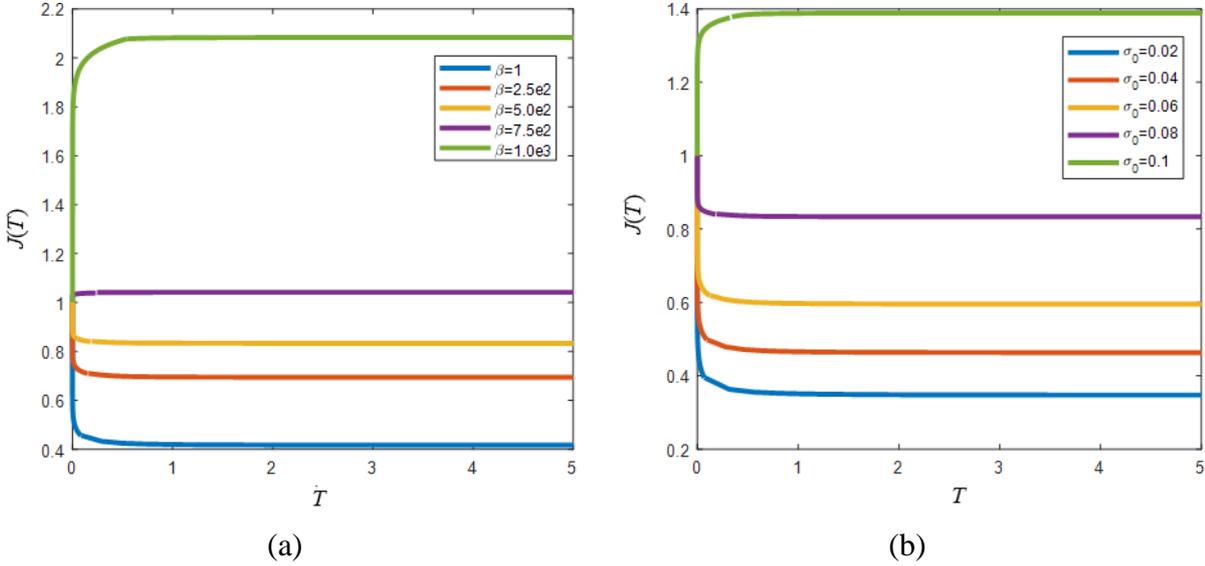

Figure 10. Well impedance versus dimensionless time for varying (a) $\beta$ and (b) $\sigma_0$.

## 5 Conclusions

Analytical modelling of the displacement of water by $CO_2$ under fines migration in layer-cake reservoir allows drawing the following conclusions.

Quasi 2D displacement problem under vertical equilibrium and fines migration in layer-cake reservoir allows for extended Buckley-Leverett solution. The model assumes hydrostatic pressure in water and gas, and domination of capillary gas-water-menisci detaching force over the attached electrostatic force. The solution is obtained in the implicit form for known absolute permeability profile k(z) and expressions for capillary pressure and relative permeability.

We show the significant effect of fines migration on water displacement – while it decreases well injectivity, fines migration simultaneously increases reservoir sweep, increasing $CO_2$ storage capacity. As the formation damage coefficient ($\beta$) increases from 1 to 1000 and the $\sigma_0$ varies from 0.01 to 0.1, sweep efficiency increases by factors of 2 and 1.5, respectively, while well injectivity decreases by factors of 5 and 3, respectively.